\documentclass[aip,jmp]{revtex4-1}

\usepackage{amssymb}
\usepackage{graphicx}
\usepackage{subfigure}
\usepackage{hyperref}
\usepackage{mathrsfs}
\usepackage{latexsym}
\usepackage{amsfonts}
\usepackage{amsmath}
\usepackage{esint}
\usepackage{amsxtra}
\usepackage[enableskew]{youngtab}
\usepackage{pifont}
\usepackage{url}

\hypersetup{
     colorlinks  = true,
     citecolor = blue,
     linkcolor = blue,
     urlcolor=blue}

\def \ep1{\epsilon_1}
\def \ep2{\epsilon_2}
\def \m{\mbox}
\def \tomega{\widetilde{\omega}}
\def \tp{\widetilde{p}}
\def \be{\begin{equation}}
\def \ee{\end{equation}}
\def\beq{\begin{eqnarray}}
\def\eeq{\end{eqnarray}}
\def \ba{\begin{array}}
\def \ea{\end{array}}
\def\nn{\nonumber}

\def \f{\frac}

\def \p{\partial}

\def \sn{\mbox{sn}}
\def \cn{\mbox{cn}}
\def \dn{\mbox{dn}}

\begin{document}

\title{Combinatorial approach to Mathieu and Lam\'{e} equations}

\author{Wei He}
\email{weihephys@gmail.com}
\affiliation{$^1$Center of Mathematical Sciences, Zhejiang University,\\
Hangzhou 310027, China\\
$^2$Instituto de F\'isica Te\'orica, Universidade Estadual Paulista,\\
Barra Funda, 01140-070, S\~{a}o Paulo, SP, Brazil}

\begin{abstract}
Based on some recent progress on a relation between four dimensional super Yang-Mills gauge theory and quantum integrable system,
we study the asymptotic spectrum of the quantum mechanical problems described by the Mathieu equation and the Lam\'{e} equation.
The large momentum asymptotic expansion of the eigenvalue is related to the instanton partition
function of supersymmetric gauge theories which can be evaluated by a combinatorial method.
The electro-magnetic duality of gauge theory indicates that in the parameter space there are three asymptotic expansions for the
eigenvalue, we confirm this fact by performing the WKB analysis in each asymptotic expansion region.
The results presented here give some new perspective on the Floquet theory about periodic differential equation.
\end{abstract}


\maketitle
\newpage
{\hypersetup{linkcolor=black}
\tableofcontents}
\section{Introduction}

Nonrelativistic quantum mechanics is naturally related to the second order differential equations.
Exactly solvable quantum mechanics models play a very fundamental role in demonstrating the basic concepts and methods of quantum theory.
A special interesting class of ordinary differential equations are those with periodic coefficients,
in quantum mechanics they describe a particle moving in one dimensional periodic potentials.
Therefore spectral problem of these equations is of particular interesting for both quantum physics and functional analysis.
In this paper we study two typical equations of this kind, the Mathieu equation and the Lam\'{e} equation.
They are respectively associated to two mathematicians who studied them in the 19th century \cite{mathieu, lame}.
The two equations and their solutions are related to each other: the Mathieu equation is a particular limit of the Lam\'{e} equation.

Our approach to the problem would be different from the conventional methods, the spectral solution of the equations are studied in the context of their relation to supersymmetric quantum gauge theory.
It is surprising that such simple quantum mechanics problems are related to seemly much more complicated quantum gauge theories.
In general the quantum dynamics of gauge theory is very complicated, nevertheless in supersymmetric gauge theory models the dynamics is tamed and many exact results are obtained. The quantum gauge theory relevant to our study is the N=2 Seiberg-Witten gauge theory \cite{SW9407, SW9408} subject to the $\Omega$ background deformation \cite{instcount}.
According to the recent proposal of Nekrasov and Shatashvili \cite{NS0901a,  NS0901b, NS0908},
the low energy dynamics of the deformed N=2 gauge theories is equivalent to the spectral problem of some quantum integrable systems.
This connection leads to two particular examples relating quantum gauge theory to the periodic Schr\"{o}dinger operator: the SU(2) pure gauge theory is related to the Mathieu equation, the SU(2) gauge theory with an adjoint matter (the N=2$^*$ theory) is related to the Lam\'{e} equation.

We summarise the structure of the paper as follows.
In Section \ref{sec2} we explain the relation between differential equations and corresponding gauge theories.
In Section \ref{sec3} we show how the asymptotic spectrum of the Mathieu equation is related to the effective action of the SU(2) pure gauge theory.
In Section \ref{sec4} we obtain the large quasimomentum asymptotic eigenvalue for the Lam\'{e} equation from the instanton partition function of the N=2$^*$ gauge theory.
In Section \ref{sec5} we perform the Wentzel-Kramers-Brillouin(WKB) analysis for the large quasimomentum eigenvalue, based on the Floquet theory, to confirm the results obtained from the gauge theory.
In Section \ref{sec6} using the same differential operators obtained in Sec. \ref{sec5}, we obtain other two eigenvalue expansions, the relation between the three eigenvalue expansions can be understood by the electro-magnetic duality of the effective gauge theory. The Section \ref{sec7} is for conclusion and open problems. Some technical details are presented in Appendixes.

\section{Differential equations and gauge theory}\label{sec2}

In this section we first explain some basic facts about the Mathieu and the Lam\'{e} equations,
in the context of Floquet theory. We are interested in their asymptotic energy spectrum which by itself is an interesting topic in quantum theory.
Then we explain how the spectral theory of  Schr\"{o}dinger operators is related to the instanton partition function of gauge theories.

\subsection{The Mathieu equation}

The Mathieu differential equation is \be
\f{d^2\Psi}{dz^2}+(\lambda-2h\cos 2z)\Psi=0.\label{matheq}\ee The
related modified Mathieu equation is obtained by $z\to iz$, then cosine potential is transformed to the hyperbolic potential.
The (modified) Mathieu equation is useful in various mathematics and
physics problems, including the separation of variables for the
wave equation in the elliptical coordinates, describing a quantum particle
moving in periodic potential. It also appears in problems such as
wave scattering by D-brane of string theory \cite{mathieuDbrane},
reheating process in some inflation models \cite{mathieucosmo}.
In quantum integrable theory it is the two body Shr\"{o}dinger equation of the Toda system.

The potential $\cos2z$ is periodic along the real axes, therefore
the Mathieu equation is a Floquet differential
equation. The Floquet (or Floquet-Bloch) solution is a function with the following monodromy property under the shift of argument by a period $\pi$,
\be \Psi(z+\pi)=e^{i\pi\nu}\Psi(z),\ee where the quantity
$\nu=\nu(\lambda,h)$ is the {\em Floquet exponent}. For periodic
Shr\"{o}dinger equation, the Floquet exponent of wave function is the quasimomenta of particles. Our focus in this paper is the asymptotic solution for the eigenvalue $\lambda$ as a function of $\nu, h$.

In general the eigenvalue relation $\lambda(\nu,h)$ can not be written in elementary functions, only
asymptotic expansions are obtained when a small expansion parameter
is available. For the Mathieu equation there are three
asymptotic expansions for $\lambda$, the leading order terms are
\be
\begin{aligned}
&\lambda\sim\nu^2+\mathcal{O}(\f{h^2}{\nu^2}),\qquad \nu\gg1, \f{h}{\nu^2}\ll1,\\
&\lambda\sim\pm2h+\mathcal{O}(\nu\sqrt{h}),\qquad \f{h}{\nu^2}\gg1.
\end{aligned}
\ee
These are well known results about the Mathieu equation,
they can be found in handbooks such as Refs.~\onlinecite{AbramowitzStegun, NIST,
MullerKirstenQUANT}. Other useful materials can be found in books Refs.~\onlinecite{McLachlan, Erdelyi, Arscott} and in papers Refs. ~\onlinecite{Mathieurefrecent0, Mathieurefrecent1, Mathieurefrecent2}.
We would explain their relation to gauge theory effective action in Sec. \ref{sec3}.

\subsection{The Lam\'{e} equation}

The Lam\'{e} differential equation contains an elliptic potential, it can be written in several forms.
In the Jacobian form it is \be
\f{d^2\Phi}{d\varkappa^2}-[A+n(n-1)k^2\sn^2\varkappa]\Phi=0,\label{lame}\ee
where $\sn\varkappa=\sn(\varkappa|k^2)$ is the Jacobi elliptic function.
In the Weierstrass form it is
\be
\f{d^2\Phi}{dz^2}-[B+n(n-1)\wp(z)]\Phi=0,\label{lame2}\ee where
$\wp(z)=\wp(z;\omega_1,\omega_2)$ is the Weierstrass elliptic
function. The coefficient $n(n-1)$ is in accordance with usual literature, but we will not
discuss whether $n$ is an integer or not which is crucial for the
classification of the solution, this does not affect the discussion of asymptotic spectrum.
The Lam\'{e} equation is obtained from separation of
variables for the Laplace equation in the ellipsoidal coordinates.
In quantum physics it is the Shr\"{o}dinger equation of two body
elliptic Calogero-Moser system.

The elliptic modulus of the function $\sn\varkappa$ is $k$, while the elliptic modulus of the function $\wp(z)$  is $q=\exp(i2\pi\f{\omega_2}{\omega_1})$,
they are related through the relation (\ref{k2ei}) given in Appendix \ref{append1}.
The Eq. (\ref{lame}) and Eq. (\ref{lame2}) are related by a change of variables \cite{Whittaker-Watson},
\be
\f{\varkappa-iK^{'}}{\sqrt{e_1-e_2}}=z,\label{cooedchange}\ee where
$K^{'}=K(k^{'})$ is the complete elliptic integrals of the first kind, $k^{'}=\sqrt{1-k^2}$ is the complementary module, and $e_i(q)=\wp(\omega_i)$.
Their eigenvalues are related by \be
B=(e_1-e_2)A-e_2n(n-1),\ee or \be
A=\f{B}{e_1-e_2}-\f{1+k^2}{3}n(n-1).\label{ABrela}\ee The two
equations are equivalent provided the change of variables is smooth.
However, the change is actually singular in the limit $k\to0$, as in
this limit $K^{'}\to\infty$. Therefore, the two equations may reduce
to different equations in limits involving $k\to0$.

The elliptic functions $\sn\varkappa$ and $\wp(z)$ are doubly periodic. The periods of $\sn\varkappa$ are $4K$ and $2iK^{'}$, the
periods of $\wp(z)$ are $2\omega_1$ and $2\omega_2$, they are
related by \be \omega_1=\f{K}{\sqrt{e_1-e_2}},\qquad
\omega_2=\f{iK^{'}}{\sqrt{e_1-e_2}}.\ee Since we have
$\sn(\varkappa+2K)=-\sn(\varkappa), \sn(\varkappa+iK^{'})=\f{1}{k\sn(\varkappa)}$,
the periods of the potential $\sn^2\varkappa$ are $2K$ and $2iK^{'}$.

The Lam\'{e} equation also falls into the equations of Floquet type, however, as the potential is the elliptic function,
a generalised version of the Floquet theory is needed.
The generalised Floquet theory for elliptic potential is one of the main results of this paper, the details are given in Secs. \ref{sec3}-\ref{sec7}.
One of the periods $2K$(or $2\omega_1$) actually can be treated by the conventional Floquet theory, see section \ref{sec5} for detail.
When the coordinate $\varkappa$ shifts a period $2K$,
or coordinate $z$ shifts a period $2\omega_1$, the phase shift of the
function $\Phi$ defines the Floquet exponent by \be
\Phi(\varkappa+2K)=e^{i2K\mu}\Phi(\varkappa),\label{phaseshlame1}\ee
or\be \Phi(z+2\omega_1)=e^{i2\omega_1\nu}\Phi(z).\ee
We use different letters $\mu$ and $\nu$ to denote the exponent for the
Lam\'{e} equation, depending on its appearance in solution of the equation in Jacobian form (\ref{lame}) or in Weierstrass form (\ref{lame2}).
The phase shifts should be the same, then we have the relation
\be
\f{\nu}{\mu}=\f{K}{\omega_1}=\sqrt{e_1-e_2}.\label{munurela}\ee In the
limit $k\to0$, $\mu$ and $\nu$ coincide. This limit is needed when
we reduce the Lam\'{e} equation to the Mathieu equation, and indeed
their eigenvalue expansions share some common features if we compare the formulae
(\ref{mathieueigen}) and (\ref{lameeigen}), (\ref{mathieueigen2}) and (\ref{lameeigen4}).

In the limit $k\to0$, we have $2K\to\pi$ and $2iK^{'}\to i\infty$, this indicates the period $2iK^{'}$ (or $2\omega_2$) is a bit different from the period $2K$(or $2\omega_1$), and indeed it needs a special treatment when concerns the Floquet theory, see section \ref{sec6}.

We would find the eigenvalue of Lam\'{e} equation, $A$ (or $B$), as a function of $\mu$ (or $\nu$) and $n, k$ (or $q$).
The Lam\'{e} eigenvalue also has three asymptotic expansion regions, their leading order terms are
\be
\begin{aligned}
&A\sim-\mu^2-\f{\kappa^2}{2}+\mathcal{O}(\f{\kappa^4}{\mu^2}),\qquad \mu\gg1, \f{\kappa}{\mu}\ll1,\\
&A\sim-i2\kappa\mu+\mathcal{O}(\mu^2),\qquad  \f{\kappa}{\mu}\gg1,\\
&A\sim-\kappa^2+i2\kappa\mu+\mathcal{O}(\mu^2),\qquad  \f{\kappa}{\mu}\gg1.\label{lameeigenapprox}
\end{aligned}
\ee
where $\kappa^2=n(n-1)k^2$. The more precise form of the expansions are in the formulae (\ref{lameeigen2}), (\ref{lameeigen4}) and (\ref{lameeigen5b}).
The first expansion was obtained recently by E. Langmann as a special case in his
solution to quantum elliptic Calogero-Moser model \cite{Langmann2004b}, however expressed in terms of $B(\nu, n, q)$.
The second expansion has been given by E. L. Ince,  and later by H. J. W.
M\"{u}ller, see the book Ref.~\onlinecite{MullerKirstenQUANT}.
To our knowledge, the third expansion is a new result not given in previous literature.
In Secs. \ref{sec4}-\ref{sec6} we derive all three asymptotic expansions in the context of the generalised version of Floquet theory.

In quantum mechanics we often study spectral solution with real eigenvalue, however, in this paper we treat all quantities $z,\lambda, h$ and $\varkappa, A, n$ (and $z, B)$ as complex variables.  The N=2 gauge theories are related to algebraic integrable systems where mechanical systems are investigated by complex analysis methods. The spectral curves of the integrable systems are defined over the field of complex numbers,
the abelian variety of the curve, which governs the linearized motion of corresponding mechanical system, is the Jacobian variety
which carries a complex structure, i.e., a complex tori. Quantization of the algebraic integrable system needs to find the eigenvalues of complex Hamiltonian operators among which is the complex Schr\"{o}dinger operator. When we compare the magnitude of these parameters with some numbers, we mean
taking their absolute values or restoring their real values.

\subsection{The spectrum of Schr\"{o}dinger operator and quantum gauge theory}\label{subsec2.3}

In this section we explain the relation between integrable systems and quantum gauge theories, which provides the basic background for our study.
It has been known that the exact solution of Seiberg-Witten on N=2 supersymmetric Yang-Mills(SYM) gauge theory can be understood by the fact that the low energy dynamics on the special Kahler moduli space of gauge theory is equivalent to the dynamics of classical algebraic integrable system\cite{integsys1, integsys2, integsys3, im9512, im9511, HokerPhong}. Classical mechanical system can be quantised, the quantization procedure can be understood as a deformation of the phase space and associated functional relation; correspondingly the gauge theory also allows a deformation by background field parameterized by $\epsilon_1,\epsilon_2$ \cite{instcount, no0306, instcount2, instcount3, instcount4}, and indeed the gauge theory deformation is exactly the quantization of classical mechanical system mentioned above. Their precise relation is recently pointed out by Nekrasov and Shatashvili \cite{NS0908}:
{\em the Coulomb vacua of $\Omega$ deformed supersymmetric gauge theories, in the limit $\epsilon_1=\hbar, \epsilon_2=0$,
are in one to one correspondence with the spectra of certain quantum integrable systems.}

The action of the gauge theory has a scalar potential which leads to the vacuum expectation value (VEV) for the scalar,  it breaks the gauge group and gives mass to some fields. The VEV takes value in the Cartan subalgebra of the gauge group, $\langle\phi\rangle=\sum_{i=1}^r\phi_iH_i$, with $r$ the rank of gauge group. The eigenvalues of the matrix $\langle\phi\rangle$ are denoted by $a_i$.
The low energy effective theory is obtained by integrating all heavy particles and instanton effects,
it is described by the prepotential function which includes a perturbative term and infinite many terms of instanton correction,
$\mathcal{F}(a_i, q_{in})=\mathcal{F}^{pert}+\sum_{\ell=1}^{\infty}\mathcal{F}_{\ell}q_{in}^\ell$,
where $q_{in}$ is the instanton expansion parameter \cite{SW9407, SW9408}. Upon the so called $\Omega$ background field deformation,
the gauge theory remains well defined, and even better: while it is very difficult to compute the instanton contribution $\mathcal{F}_{\ell}(a_i,q_{in})$ in the undeformed gauge theory by quantum field theory method, the deformed prepotential $\mathcal{F}(a_i, q_{in}, \epsilon_1, \epsilon_2)$ can be exactly computed by a localization method \cite{instcount}. In the limit $\epsilon_2=0$, the gauge theory is effectively confined on a two dimensional plane,
therefore the Gauge/Bethe correspondence of Ref.~\onlinecite{NS0901a, NS0901b} relates the $\epsilon_1$ deformed gauge theory to quantum integrable system. More precisely, $\epsilon_1$ in the gauge theory plays the role of Plank constant, the $a_i$ are identified with the
quasimomenta of excitations in integrable model, and the vacuum expectation value
of gauge invariant operators $\langle\m{Tr}\phi^\ell\rangle$ are identified with
eigenvalues of the Hamiltonian operators $H_\ell$. Finally, the prepotential of gauge theory is identified
with the Yang-Yang potential, the equation selecting the vacua of gauge theory is the Bethe equation selecting the allowed quasimomenta of integrable model.

In a few simple cases this correspondence relates SU(2) gauge theories to 2-body quantum mechanics problems, provides simple examples to examine the relation in detail. The SU(2) pure gauge theory is related to the Schr\"{o}dinger equation of 2-particle periodic Toda model, i.e. the Mathieu equation;
the SU(2) gauge theory with adjoint matter is related to the Schr\"{o}dinger equation of 2-particle periodic elliptic Calogero-Moser model, i.e. the Lam\'{e} equation \cite{NS0908}.
The eigenvalues of Mathieu and Lam\'{e} equations are generally transcendental,
in this paper we are interested in their asymptotic solutions. Some results about their asymptotic solutions are already known,
here we revise the problem from the perspective of quantum gauge theory.
Our focus is the asymptotic eigenvalue, we do not try to solve the asymptotic wave function, although it is argued to be related to gauge theory partition function in the presence of surface operator, and can be computed by localization \cite{at1005, mt1006}.

For the SU(2) gauge theory, the only nontrivial $\m{Tr}\phi^\ell$ is for $\ell=2$, it gives the eigenvalue for the Schr\"{o}dinger equation.
The expectation value $\langle\m{Tr}\phi^2\rangle$ is a function of gauge theory parameters,
\be \langle\m{Tr}\phi^2\rangle=u(a, q_{in}, \epsilon_1, \epsilon_2),\ee
The Matone's relation \cite{matone1, matone2} states a relation between $u$ and the prepotential $\mathcal {F}$,
\be 2u=q_{in}\f{\p}{\p
q_{in}}\mathcal{F}.\label{matonerel0}\ee  As $u$ also expands according to the instanton parameter, $u=\sum_{\ell=0}^{\infty}u_{\ell}q_{in}^\ell$,
therefore we have
\be 2u_0=q_{in}\f{\p}{\p q_{in}}\mathcal {F}^{pert},\qquad 2u_\ell=\ell\mathcal {F}_\ell,\qquad
\ell\geqslant1.\label{matonerel}\ee
This relation holds for generic deformation $\epsilon_{1,2}$, after taking the limit $\epsilon_2=0$ we can interpret the relation in the context of quantum mechanical problem. As a consequence, if we properly identify a parameter in the quantum mechanical problem with the gauge theory parameter $q_{in}$,
then the $\ell$-th coefficients of $q_{in}$-expansion of the eigenvalue is proportional to
the $\ell$-th instanton contribution of the SU(2) gauge theory prepotential.

\section{Spectrum of the Mathieu equation}\label{sec3}

As the Mathieu and the Lam\'{e} equations are closely related to each
other, the procedures of deriving their eigenvalues are largely
parallel. The results in this section have already been given in
earlier works \cite{FP0208, MM0910, wh1006a, wh1006b, wh1103}. Here we give
a brief summary on the Mathieu eigenvalue as an educational example to
illustrate, and to further explain,  the basic method and logic of our approach.

There are three aspects we would explain. (1) One of the asymptotic
eigenvalue, for which $\nu\gg1$,  is directly related the gauge
theory partition function, in the way explained in subsection
\ref{subsec2.3}. (2) An independent exact WKB analysis can be
performed for the Schr\"{o}dinger equation with periodic potentials
to obtain the asymptotic eigenvalue for $\nu\gg1$. During the
process we obtain a tower of higher order differential operators.
(3) Lastly, the electro-magnetic duality of gauge theory indicates
other two asymptotic expansions for the eigenvalue. Using the
differential operators obtained before we can compute the other two
asymptotic expansions.

After understanding the procedure for the Mathieu eigenvalue, it is straightforward to apply this method to the Lam\'{e} eigenvalue problem.

\subsection{Combinatorial evaluating in the electric region}

\begin{figure}
\begin{minipage}[t]{0.5\linewidth}
\centering
\includegraphics[width=5cm]{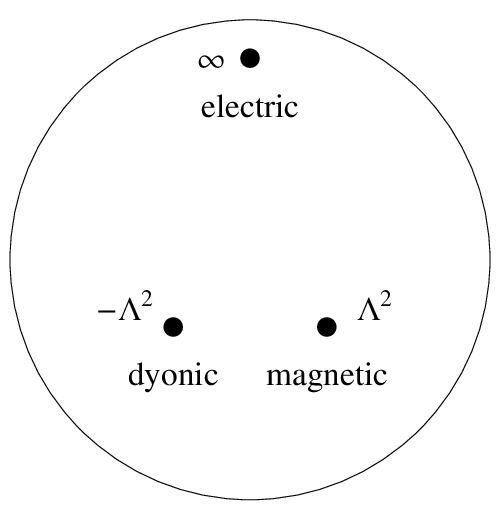}
\caption{Singularities in the $u$-plane.} \label{singpure}
\end{minipage}%
\begin{minipage}[t]{0.5\linewidth}
\centering
\includegraphics[width=8cm]{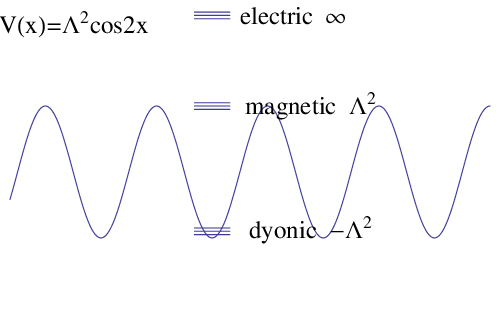}
\caption{Spectrum and duality of cosine-potential.}
\label{mathieuspectrum}
\end{minipage}
\end{figure}

First we state the relation between singularities in the moduli space
of the gauge theory and asymptotic expansion regions of the Mathieu
eigenvalue. The moduli space of the N=2 SU(2) pure
gauge theory is parameterized by the only complex gauge invariant variable $u=\langle\m{Tr}\phi^2\rangle$ which breaks SU(2)
symmetry to abelian U(1). The moduli space is singular at
$u=\infty, \pm\Lambda^2$ in the sense that at each singularity there are massless particles
that can be used to describe the effective gauge theory as a weakly coupled theory.
These singularities are labeled by the U(1) charges of the
massless particles, denoted as ``electric'', ``magnetic'' and
``dyonic'', respectively. Near each singularity the gauge theory has
an unique weak coupling description, therefore perturbative
expansion is valid. Our calculation in Refs.~\onlinecite{wh1006a, wh1006b} shows that
there is a one to one correspondence between the perturbative
expansion of gauge theory near a singularity in the  moduli space
and the asymptotic expansion for the corresponding Mathieu
eigenvalue. In each asymptotic expansion region the
intervals between adjacent energy levels are very small compared to
the eigenvalues themselves, therefore the eigenstates are dense
there, see Fig. \ref{singpure} and Fig. \ref{mathieuspectrum}.

The N=2 gauge theory is formulated by a classical Lagrangian suitable for weak coupling region near $u\sim\infty$ where $\f{\Lambda}{a}\ll1$,
the combinatorial method is used to compute the effective prepotential $\mathcal{F}(a, q_{in}, \epsilon_1, \epsilon_2)$ at that point \cite{instcount}.
Along with the logic explained in Sec. \ref{sec2}, we identify the parameters of gauge
theory and Mathieu eigenvalue as \cite{wh1006a}\be
\lambda=\f{8u}{\epsilon_1^2},\qquad
\nu=\f{2a}{\epsilon_1},\qquad
h=\f{4\Lambda^2}{\epsilon_1^2}=\f{4\sqrt{q_{in}}}{\epsilon_1^2}.\label{indent1}\ee

In order to compute the formula (\ref{matonerel0}), we need to compute the Nekrasov partition function from which the deformed prepotential $\mathcal {F}(a, q_{in}, \epsilon_1, \epsilon_2)$ can be derived \cite{instcount}.
There are formulae convenient for program treatment \cite{FP0208, bfmt}, the complete expressions of the first three $\mathcal{F}_1,
\mathcal{F}_2, \mathcal{F}_3$ are presented in Ref.~\onlinecite{FP0208}. As
observed in Ref.~\onlinecite{wh1103}, according to the Matone's relation
(\ref{matonerel0}) , and taking into account (\ref{indent1}), the
Mathieu eigenvalue is\beq
\lambda&=&\nu^2+\sum_{\ell=1}^{\infty}\f{\epsilon_1^{4\ell-2}}{16^\ell}4\ell\mathcal{F}_\ell(2a,\epsilon_1,\epsilon_2=0)h^{2\ell}\nn\\
&=&\nu^2+\sum_{\ell=1}^{\infty}4\ell\mathcal{F}_\ell(\nu,1,0)(\f{h}{4})^{2\ell}.\eeq
The leading term $\nu^2$ comes from the perturbative contribution.
We have used the fact that the numerator and denominator of
$\mathcal{F}_\ell(a,\epsilon_1)$ are homogeneous polynomials of $a, \epsilon_1$ of degrees $2\ell+2$ and $6\ell$, respectively.
We get the asymptotic expansion
\be
\lambda=\nu^2+\f{1}{2(\nu^2-1)}h^2+\f{5\nu^2+7}{32(\nu^2-1)^3(\nu^2-4)}h^4+\cdots,\label{mathieueigen}\ee
for $h\ll1, \nu\gg1$ and/or $h\gg1, \f{h}{\nu^2}\ll1$.

\subsection{Extension to other regions}

A crucial ingredient of the  Seiberg-Witten theory \cite{SW9407} is that the effective gauge theory demonstrates electro-magnetic duality in the moduli space,
the dual descriptions are valid at $u=\pm\Lambda^2$, accordingly, this indicates there are other two asymptotic expansions for the eigenvalue.
In order to extend the asymptotic expansion from $u\sim\infty$ to other two regions near
$u=\pm\Lambda^2$, we recall how the duality is studied in the undeformed gauge theory where $\epsilon_1=\epsilon_2=0$.
Actually the situation at $-\Lambda^2$ is related to that at $\Lambda^2$ by a simple map, therefore we can simplify the following explanation a bit by focusing on the singularity at $\Lambda^2$.

In order to determine the prepotential, the dual scalar VEV $a_D$ is defined and related to $a$ by the relation $a_D=\p_a\mathcal{F}$.
Moreover, $a$ and $a_D$ are given by integrals of the Seiberg-Witten form along the homology circles $\alpha$ and $\beta$ of the
Seiberg-Witten curve determined by the moduli space which is a torus for SU(2) gauge theory, $a=\oint_{\alpha}\lambda_{SW}$ and $a_D=\oint_{\beta}\lambda_{SW}$. As $\lambda_{SW}$ depends on the parameters $u, \Lambda^2$,
in this way we get relations $a(u, \Lambda^2)$ and $a_D(u, \Lambda^2)$. The two functions are globally defined on the moduli space,
$a(u, \Lambda^2)$ has an asymptotic expansion at $u\sim\infty$, $a_D(u, \Lambda^2)$ has an asymptotic expansion at $u=\Lambda^2$.

The gauge theory with $\epsilon_1\ne0,\epsilon_2=0$ is the ``quantized''
version of theory with $\epsilon_1=\epsilon_2=0$. As first studied in Ref.~\onlinecite{MM0910}, there is a way to incorporate the $\epsilon_1$ deformation into the integrand $\lambda_{SW}$, therefore we have the quantized Seiberg-Witten form $\lambda_{SW}(\epsilon_1)$.
The integrations of $\lambda_{SW}(\epsilon_1)$ along the homology circles of the torus give us the deformed relation $a(u,\Lambda^2,\epsilon_1)$ and $a_D(u,\Lambda^2, \epsilon_1)$.
The relation $a(u,\Lambda^2,\epsilon_1)$ also has an asymptotic expansion at $u\sim\infty$, by the identification (\ref{indent1}),
the inverse series of the expansion is exactly the eigenvalue solution (\ref{mathieueigen}).
At the point $u=\Lambda^2$, the dual relation $a_D(u,\Lambda^2, \epsilon_1)$ has an asymptotic expansion, it turns out that the inverse series of the expansion gives another eigenvalue solution (\ref{mathieueigen2}).

The remaining detail is how to compute the deformed integrand $\lambda_{SW}(\epsilon_1)$.
As the deformed gauge theory is related to certain quantum mechanics problem,
it is not surprising that the WKB method provides an efficient method to obtain higher order
quantum corrections. Actually in Ref.~\onlinecite{MM0910} it shows that $\lambda_{SW}(\epsilon_1)$ can be
generated from the undeformed $\lambda_{SW}$ by the action of certain differential
operators with respect to $u$ and $\Lambda^2$. In practice, setting
$\Lambda^2=1$ will further simplify the procedure, it can be restored by dimensional consideration.
We use $D(u,\p_u,\epsilon_1)$ to represent the differential operator, then the deformed relation $a(u,\epsilon_1)$ is computed by
 \be
a(u,\epsilon_1)=D(u,\p_u,\epsilon_1)\oint_{\alpha}\lambda_{SW}.\label{alphaint}\ee
The integral $a(u)=\oint_\alpha\lambda_{SW}$ is the leading order WKB approximation, higher order $\epsilon_1$ corrections are generated by $D(u,\p_u,\epsilon_1)$. The operator $D(u,\p_u,\epsilon_1)$ can be expanded as
$D(u,\p_u,\epsilon_1)=1+\sum_{n=1}^{\infty}\epsilon_1^n D_n(u,\p_u)$, here the $n$ should not be confused with the $n(n-1)$ in the Lam\'{e} potential. The differential operators $D_n(u,\p_u)$ are polynomials of $u$ and
$\p_u$, the first few of them can be found in Refs.~\onlinecite{wh1006a, wh1006b}.
Restoring the $\Lambda^2$ dependence and reversing the large $u$ asymptotic series of $a(u,\Lambda^2,\epsilon_1)$, we get the series $u(a,\Lambda^2,\epsilon_1)$ which gives the eigenvalue (\ref{mathieueigen}) through parameters identification (\ref{indent1}).

The dual relation $a_D(u,\epsilon_1)$ is given by integral of the same deformed integrand along the dual homology
circle,\be
a_D(u,\epsilon_1)=D(u,\p_u,\epsilon_1)\oint_{\beta}\lambda_{SW}.\label{betaint}\ee
After restoring the dependence on $\Lambda^2$, the function $a_D(u,\Lambda^2,\epsilon_1)$ has an asymptotic series at $u=\Lambda^2$ and its reverse series is $u=(a_D,\Lambda^2,\epsilon_1)$. In order to relate it to the eigenvalue solution of the Mathieu equation, we identify the parameters as
\be
\lambda=\f{8u}{\epsilon_1^2},\qquad
\nu=\f{2ia_{D}}{\epsilon_1},\qquad
h=\f{4\Lambda^2}{\epsilon_1^2}=\f{4\sqrt{q_{in}}}{\epsilon_1^2}.\label{indent2}\ee
Then the asymptotic series expansion of $u(a_D,\Lambda^2,\epsilon_1)$
gives another well known eigenvalue expansion, \be
\lambda=2h-4\nu\sqrt{h}+\f{4\nu^2-1}{2^3}+\f{4\nu^3-3\nu}{2^6\sqrt{h}}+\cdots,\label{mathieueigen2}\ee
for $h\gg1, \f{h}{\nu^2}\gg1$. The details are given in Ref.~\onlinecite{wh1006b}, see there for result up to the
order $h^{-\f{7}{2}}$. Moreover, the third expansion region near
$u=-\Lambda^2$, the dyonic region, is mirror of the region near
$u=\Lambda^2$. The eigenvalue there is obtained from (\ref{mathieueigen2}) by the map $\nu\to
i\nu, h\to-h$, \be
\lambda=-2h+4\nu\sqrt{h}-\f{4\nu^2+1}{2^3}-\f{4\nu^3+3\nu}{2^6\sqrt{h}}+\cdots.\label{mathieueigen22}\ee

In Appendix \ref{append4}, we give an algorithm that gives the operator $D(u,\p_u,\epsilon_1)$, therefore the deformed $\lambda_{SW}(\epsilon_1)$, up to arbitrary higher order of $\epsilon_1$.

\section{Combinatorial approach to the Lam\'{e} eigenvalue}\label{sec4}

We adopt the same method to the Lam\'{e} equation which is quite similar to the Mathieu equation. In
the limit $n\to\infty, k\to0$ while keep $n(n-1)k^2=\kappa^2$ fixed,
the Jacobi $sn$-function becomes the trigonometric function $sin$,
and Eq. (\ref{lame}) becomes equivalent to Eq. (\ref{matheq}). According to the argument of quantum field theory,
the moduli space of the corresponding N=2$^*$ gauge theory also has three singularities, hence there are three asymptotic expansions for the Lam\'{e} eigenvalue.
In this section we derive the eigenvalue expansion at the electric region from the deformed prepotential of gauge theory, and write it in a proper form.

\subsection{Identify parameters}

Firstly, we have to make the identification between the module
parameters in the Lam\'{e} equation and in the N=2$^*$
gauge theory. The parameter $k$ appearing in Eq. (\ref{lame}) is the
elliptic module of the Jacobi elliptic function. The Weierstrass elliptic function uses another module parameter, the nome $q$.
The two parameters are related to each other, their precise relation is
\be
k^2=16q^{\f{1}{2}}-128q+704q^{\f{3}{2}}-3072q^2+\cdots=\f{\theta_2(q)^4}{\theta_3(q)^4}.\label{kqrelat}\ee
Our claim is that $q$ is exactly the instanton expansion parameter
of the N=2$^*$ theory, and is the ``good'' expansion
parameter for the eigenvalue $B$ in the electric region; while $k$ is the ``good'' module
parameter for the eigenvalue $A$ in the magnetic and dyonic regions.

Another thing we have to identify is the relation between the
parameter $n$ in the Lam\'{e} equation and the adjoint mass of the
gauge theory. The equivalent parameter $m$ appears in the instanton
partition function of N=2$^*$ theory, as clarified in Ref.~\onlinecite{op1004}, the parameter $m$ differs from the
physical mass $m^*$ by $\epsilon_1, \epsilon_2$ shift,
$m=m^*+(\epsilon_1+\epsilon_2)/2$. We identify $n=\epsilon_1^{-1}m$.

Lastly, we have to identify the relation between the eigenvalue $A$,
or $B$, and the scalar condensation $u$. The Seiberg-Witten curve of
the undeformed N=2$^*$ theory is \be
y^2=(x-e_1\tilde{u}-\f{1}{8}e_1^2m^2)(x-e_2\tilde{u}-\f{1}{8}e_2^2m^2)(x-e_3\tilde{u}-\f{1}{8}e_3^2m^2),\label{swcurveadjoint}\ee
with $\tilde{u}$ related to the VEV of scalar field by \cite{dkm96, fmpt05}
\be
u=\langle\m{Tr}\phi^2\rangle=\tilde{u}-\f{m^2}{24}(1-2E_2),\label{utildeu1}\ee where $E_2(q_{in})$ is the
second Eisenstein series. We have rescaled the mass in Ref.~\onlinecite{SW9408} by $m^2\to\f{m^2}{2}$ in
order to match the mass parameter that appears in the instanton
counting formula. Again this is an elliptic
curve, with two conjugate homology circles $\alpha$ and $\beta$, and
the Seiberg-Witten differential one form. $\tilde{u}$ and $u$ differ
by terms caused by the mass deformation. It turns out that $\tilde{u}$
is directly related to the eigenvalue $B$.

In summary, we identify \be B=-\f{8\tilde{u}}{\epsilon_1^2},\quad
\nu=\f{2a}{\epsilon_1},\quad n=\f{m}{\epsilon_1}, \quad q=q_{in}.\label{indent3}\ee

\begin{figure}[h]
  \centering
  \includegraphics[width=5cm]{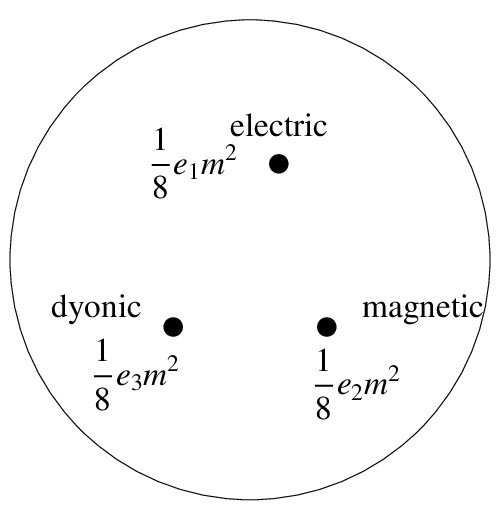}
   \caption{Singularities in the $\tilde{u}$-plane for N=2$^*$ theory.}
\label{fig3}
\end{figure}

The moduli space has three singularities in the $\tilde{u}$ plane,
as showed in Fig. \ref{fig3}, they are located at $\tilde{u}=\f{1}{8}e_im^2, i=1,2,3$.
The elliptic curve (\ref{swcurveadjoint}) degenerates at each singularity.
Note that the stationary points of the potential $\wp(z)$ are
exactly at $z_{1,2,3}$ where $\p_z\wp(z)|_{z_i}=0,
\p_z^2\wp(z)|_{z_i}\ne0$ and $\wp(z_i)=e_i$. In the $u$ plane, the singularities are located at
\be
\begin{aligned}
&u=m^2(\f{1}{8}-4q^2-12q^4+12q^5+\cdots),\\
&u=\mp m^2(\sqrt{q}\pm3q+4q^{\f{3}{2}}\pm7q^2+\cdots).
\end{aligned}\label{singularityinu}
\ee
As in the
case for the pure SU(2) SYM (the Mathieu equation), there are only three
singularities in the moduli space, therefore, there are only three
asymptotic expansion regions for the eigenvalue. While the Mathieu
eigenvalue has two free parameters $h$ and $\nu$, the Lam\'{e}
eigenvalue has three free parameters, $n,k$ and $\nu$, it is not a
prior fact the number of expansion regions remains to be three.
However, the gauge theory argument indicates there is no other expansion
region.

\subsection{Combinatorial evaluating in the electric region}\label{subsec4.2}

We derive the eigenvalue in the electric region. Let us look at the
perturbative part, in the limit $\epsilon_2=0$, the deformed prepotential is
given by \cite{NS0908} \be \mathcal{F}^{pert}=\pi
i\tau\sum_{i=1}^{2}a_i^2+2\pi i\epsilon_1\sum_{i,j=1}^{2}
(\varpi_{\epsilon_1}(a_{ij})-\varpi_{\epsilon_1}(a_{ij}-m-\epsilon_1)),\ee
due to the convention of instanton computation the variable $a$ for SU(2) gauge theory is represented by  $a_1,a_2$ satisfying $a_1=-a_2=a$, while $a_{ij}=a_i-a_j$ and the complex gauge coupling constant $\tau$ is related to the instanton expansion parameter by $q_{in}=\exp(2\pi i\tau)$, the definition of the function $\varpi_{\epsilon_1}(x)$ can be found in Ref.~\onlinecite{NS0908}. Only the first term $a^2\ln q_{in}$ contributes to the eigenvalue.

Then we come to the instanton part of the gauge theory, or the $q$
correction part of the Lam\'{e} eigenvalue. Following the line of
instanton counting, we derive the prepotential $\mathcal
{F}_\ell(a_{12}, m, \epsilon_1,\epsilon_2), \ell=1,2,3,\cdots$, and set
$\epsilon_2=0$, the eigenvalue $B$ can be  obtained via the Matone's
relation. Without presenting the results of $Z_\ell, \ell=1,2,3\cdots$, which are
fairly lengthy, we only give the final results of the prepotential.
For example, the 1-instanton contribution is \be
\mathcal{F}_1q_{in}=\f{2m(m-\epsilon_1)[m(m-\epsilon_1)-a_{12}^2+\epsilon_1^2]}{a_{12}^2-\epsilon_1^2}q_{in}.\ee
Writing in terms of $n,\nu,q$, it is \be
\mathcal{F}_1q_{in}=\epsilon_1^2\f{2n(n-1)q}{\nu^2-1}\times[n(n-1)-(\nu^2-1)].\ee
In a similar way we can compute all other coefficients $\mathcal{F}_\ell$.  A notable fact is that in
the final results of instanton counting the parameter $n$ always
appears in the form $n(n-1)$, consistent with its appearance in the
Lam\'{e} equation. This is already visible in the numerator of counting formula in Ref.~\onlinecite{bfmt} where terms involving $m$ combine to $m(m-\epsilon_1)$.

According to the Matone's relation, the scalar condensation $u$ is
given by $2u=q\f{\p}{\p q}\mathcal{F}$, and in the presence of $\epsilon_1$ deformation,  $\tilde{u}$ and $u$ are related by\be
\tilde{u}=u+\f{m(m-\epsilon_1)}{24}(1-2E_2(q)),\ee here we have
turned on the $\epsilon_1$ quantum correction to the classical
relation (\ref{utildeu1}), this is the only consistent way to deform
the classical relation. Now we are in the position to get the
eigenvalue $B$, using identification (\ref{indent3}),
\beq B&=&-\f{4}{\epsilon_1^2}q\f{\p}{\p q}\mathcal{F}-\f{n(n-1)}{3}(1-2E_2(q))\nn\\
&=&-\nu^2-\f{n(n-1)}{3}(1-2E_2(q))-\f{8n(n-1)q}{\nu^2-1}[n(n-1)-(\nu^2-1)]\nn\\
&\quad&-\f{8n(n-1)q^2}{(\nu^2-1)^3(\nu^2-4)}\times[
n^3(n-1)^3(5\nu^2+7)-12n^2(n-1)^2(\nu^2-1)^2\nn\\
&\quad&+6n(n-1)(\nu^2-1)^2(\nu^2-2)-3(\nu^2-1)^3(\nu^2-4)]+\mathcal{O}(q^3).\label{lameeigen}\eeq
The term $-\nu^2$ comes from the perturbative result $q\p_q\mathcal
{F}^{pert}=\epsilon_1^2\nu^2/4$. The expansion (\ref{lameeigen}) for $B$ is valid when
\be q\ll1,\quad n\gg1,\quad \nu\gg1,\quad
nq^{\f{1}{4}}\ll\nu.\ee  Close to the electric singularity, we have the
leading order behavior $8u\sim m^2$, that is $\nu^2\sim n(n-1)$,
therefore $B\sim-\f{2}{3}n(n-1)+\mathcal {O}(n(n-1)q)$. An equivalent
expansion was also obtained by E. Langmann in Ref.~\onlinecite{Langmann2004b} by a different method.
We have shown the equivalence of his
eigenvalue expansion $\mathcal {E}$ and our expansion $B$, see Appendix \ref{append3}.

At this point we should emphasise a subtle point, as explained in Ref.~\onlinecite{agt}, the Nekrasov partition function actually computes the instanton effects of U(2) gauge theory, even after setting $a_1=-a_2=a$ due to the traceless condition of SU(2),  it differs from the SU(2) theory by an U(1) factor.
Therefore the deformed prepotential $\mathcal{F}$ in this subsection should be understood as U(2) prepotential subject to the traceless condition,
it differs from the SU(2) prepotential by the value $\f{1}{12}m(m-\epsilon_1)(1-E_2)$. There is no such U(1) factor difference for pure gauge theories with gauge groups U(2) and SU(2).

In order to compare (\ref{lameeigen})  with the WKB results, we rewrite the asymptotic expansion in terms of $A(\mu,k)$.
It can be derived from the expansion of $B$, taking into
account the relation between $A$ and $B$ in (\ref{ABrela}), the
relation between $\mu$ and $\nu$ in (\ref{munurela}), and the relation between $k$ and $q$ in (\ref{kqrelat}).  We get
\beq A&=&-\mu^2-\f{1}{2}n(n-1)k^2-n(n-1)(\f{1}{16}k^4+\f{1}{32}k^6+\f{41}{2048}k^8+\f{59}{4096}k^{10}+\cdots)\nn\\
&\quad&-\f{n^2(n-1)^2}{\mu^2}[\f{1}{32}k^4-\f{1}{4096}k^8-\f{1}{4096}k^{10}+\cdots]\nn\\
&\quad&-\f{n^2(n-1)^2}{\mu^4}[\f{1}{32}k^4-\f{1}{64}k^6-(\f{7+6n(n-1)}{4096})k^8-(\f{7+6n(n-1)}{8192})k^{10}+\cdots]\nn\\
&\quad&+\mathcal{O}(\f{1}{\mu^6}).\label{lameeigen2}\eeq

Before doing the WKB check, we first give a simple consistent check
using a limit of the Lam\'{e} equation. The N=2$^*$
theory has a decoupling limit. When the energy scale is very small,
or the mass is very large, the adjoint mass decouples and the theory
flows to the pure gauge theory. The limit is $q\to0, m\to\infty$ with $m^2\sqrt{q}\to-\Lambda^2$;
or equivalently $k\to0, m\to\infty$, with $m^2k^2\to-16\Lambda^2$.
In this limit, every singularity in the moduli space of the N=2$^*$ theory becomes a singularity of the pure gauge theory.
This can be clearly seen in (\ref{singularityinu}).
Writing in terms of the parameters of the differential equation, the limit is $q\to0, n\to\infty$ with $n(n-1)\sqrt{q}\to-\f{h}{4}$;
or equivalently $ k\to0, n\to\infty$ with $n(n-1)k^2\to-4h$.  Accordingly, in this
limit the Lam\'{e} equation reduces to the Mathieu equation, and the
Lam\'{e} eigenvalue expansions (\ref{lameeigen}) and (\ref{lameeigen2}) reduce
to the Mathieu eigenvalue expansion (\ref{mathieueigen}), $-B\to\lambda-\f{1}{3}n(n-1),  -A\to\lambda-2h$.

\section{Eigenvalue from the WKB analysis}\label{sec5}

We perform the WKB perturbation to confirm that the asymptotic expansions in (\ref{lameeigen}) and (\ref{lameeigen2}) are indeed eigenvalue solutions for the Lam\'{e} equation. The method is explained in Section \ref{sec3}, for some periodic potentials including the elliptic potential the higher order WKB perturbations are generated from the leading order by the action of differential operators \cite{MM0910}.
In this section, we do the same thing for the Lam\'{e} equation, as
have done for the Mathieu equation in Refs.~\onlinecite{MM0910, wh1006a, wh1006b}.

Equation (\ref{lame}) is rewritten as \be
\f{\epsilon^2}{2}\Phi^{''}-(\omega+\sn^2\varkappa)\Phi=0,\label{lame3}\ee
where \be \epsilon^2=\f{2}{n(n-1)k^2},\qquad
\omega=\f{A}{n(n-1)k^2}.\ee Note that $\epsilon_1$ is contained in $\epsilon$. Suppose $\epsilon\ll1$, i.e.,
$nq^{\f{1}{4}}\gg1$, then $\epsilon$ is a small parameter for the WKB expansion. We choose Eq. (\ref{lame}) to do the WKB analysis because
there are readily available integral formulae for the Jacobi elliptic functions. It is legal to choose Eq.
(\ref{lame2}) instead, as in Refs.~\onlinecite{fl0912, mt1006} where the leading
order expansion is performed. Expanding the function $\Phi$ as WKB
series, \be \Phi(\varkappa)=\exp
i\int_{\varkappa_0}^{\varkappa}d\varkappa^{'}(\f{p_0(\varkappa^{'})}{\epsilon}+p_1(\varkappa^{'})+\epsilon
p_2(\varkappa^{'})+\epsilon^2 p_3(\varkappa^{'})+\cdots),\ee then substitute into the equation,  $p_n$ can be determined recursively,
\be
\begin{aligned}
&p_0=i\sqrt{2}\sqrt{\omega+\sn^2\varkappa},\qquad
p_1=\f{i}{2}(\ln(p_0))^{'},\qquad p_2=\f{1}{4p_0}[\f{3}{2}(\f{p_0^{'}}{p_0})^2-\f{p_0^{''}}{p_0}],\\
&p_3=\f{i}{2}(\f{p_2}{p_0})^{'},\qquad p_4=\f{i}{2}(\f{p_3}{p_0})^{'}-\f{p_2^2}{2p_0},\qquad \cdots\label{piwkb}
\end{aligned}
\ee
where the prime denotes $\f{\p}{\p\varkappa}$.

According to the Floquet theory, in order to obtain the dispersion relation $A(\mu)$ we have to evaluate the contour integral $\oint_\alpha pd\varkappa$, where $\alpha$ is the homology cycle of the curve
(\ref{swcurveadjoint}) relevant to electric scalar VEV. Let us first carry out the integral for the leading order integrand $\oint_\alpha p_0d\varkappa$.
The integral can be computed using the {\em amplitude}, defined by $ \varphi=\m{am}\varkappa$,  then
we have \be \sn\varkappa=\sin\varphi,\qquad
d\varkappa=\f{d\varphi}{\sqrt{1-k^2\sin^2\varphi}}.\ee The integral
becomes \be
\oint_\alpha\sqrt{\omega+\sn^2\varkappa}d\varkappa=2\int_{0}^{K}\sqrt{\omega+\sn^2\varkappa}d\varkappa
=2\int_{0}^{\f{\pi}{2}}\sqrt{\f{\omega+\sin^2\varphi}{1-k^2\sin^2\varphi}}d\varphi.
\ee In the electric region, $A\gg\kappa^2$, therefore $\omega\gg1$,
we have the expansion \be
\sqrt{\omega+\sin^2\varphi}=\sqrt{\omega}(1+\f{\sin^2\varphi}{2\omega}-\f{\sin^4\varphi}{8\omega^2}+\f{\sin^6\varphi}{16\omega^3}
+\cdots).\ee All the integrals we
need to do are of the form\be
\int_{0}^{\f{\pi}{2}}\f{\sin^{2n}\varphi}{\Delta}d\varphi,\qquad\m{with}\quad
\Delta=\sqrt{1-k^2\sin^2\varphi},\qquad n=0,1,2,3,\cdots\ee They are
given in Chapter 2 of Ref.~\onlinecite{tableofintegrals}. After collecting all
terms together, we get
\beq
\oint_\alpha\sqrt{\omega+\sn^2\varkappa}d\varkappa
&=&2K\sqrt{\omega}+\f{K-E}{k^2}\omega^{-1/2}-\f{(2+k^2)K-2(1+k^2)E}{12k^4}\omega^{-3/2}\nn\\
&\quad&+\f{(8+3k^2+4k^4)K-(8+7k^2+8k^4)E}{120k^6}\omega^{-5/2}+\cdots,\label{p0lame}\eeq where $K$ and $E$ are the complete
elliptic integrals of the first and second kind, respectively.

For higher order contour integrals, similar to the case of Mathieu equation, they can be obtained from higher order differential operators acting on $\oint_\alpha p_0d\varkappa$,
\be \oint_\alpha p_nd\varkappa=D_n(\omega,\p_\omega,k)\oint_\alpha p_0d\varkappa.\ee
The odd order integrals actually vanish because the integrand are total derivatives, \be
\oint_{\alpha}p_{2l+1}d\varkappa=0,\qquad l=0,1,2\cdots\ee
Hence we can set all odd differential operators $D_{2l+1}(\omega,\p_\omega,k)=0$.
While for the even order integrals, using the trick of Ref.~\onlinecite{MM0910}, we get the operator
$D_{2l}(\omega,\p_\omega,k)$. For example, we get $D_2(\omega,\p_\omega,k)$ from
\beq \oint_{\alpha}p_2d\varkappa&=&-\f{1}{12}[(1+2\omega+2k^2\omega
+3k^2\omega^2)\p_\omega^2\nn\\
&\quad&+(1+k^2+3k^2\omega)\p_\omega-\f{3}{4}k^2]\oint_{\alpha}p_0d\varkappa.\label{p2lame}\eeq
And $D_4(\omega,\p_\omega,k)$ is given by
\beq \oint_{\alpha} p_4d\varkappa&=&\f{1}{64}\lbrace\f{2}{135}[21+(84+359k^2)\omega+(84+1394k^2+359 k^4)\omega^2\nn\\
&\quad&+k^2(1077+1352k^2)\omega^3+1014k^4\omega^4]\p^4_\omega\nn\\
&\quad&+\f{4}{27}[(18+73k^2)+(36+341k^2+146k^4)\omega+k^2(432+597k^2)\omega^2+597k^4\omega^3]\p^3_\omega\nn\\
&\quad&+\f{1}{18}[(60+191k^2+225k^4)+k^2(667+1162k^2)\omega+1743 k^4\omega^2]\p^2_\omega\nn\\
&\quad&+\f{1}{18}[k^2(8+63k^2)+189k^4\omega]\p_\omega-\f{k^4}{8}\rbrace\oint_{\alpha} p_0d\varkappa.\label{p4lamemini}\eeq
In  Appendix \ref{append4},
we give a systematic method to obtain higher order differential operators.

The contour integral along the $\alpha$ circle computes the phase
shift of the function $\Phi(\varkappa)$ in (\ref{phaseshlame1}), we have
\be 2K\mu=\oint_\alpha d\varkappa
p(\varkappa)=(1+\sum_{n=1}^{\infty}\epsilon^nD_n(\omega,\p_\omega,k))\oint_\alpha
d\varkappa\f{p_0(\varkappa)}{\epsilon}.\ee Therefore the Floquet
exponent is given by\be \mu=\f{1}{2K}\oint_\alpha d\varkappa
p(\varkappa).\ee

The remaining work is straightforward but tedious: expand the $p_0$
integral (\ref{p0lame}) as power series of $\omega$ and $k$, use the
differential operators $D_{2l}(\omega,\p_\omega,k)$ to
generate higher order integrals for $p_2, p_4,\cdots$, we get the series
expansion of $\mu=\mu(\omega,k)$. Then inverse the series
$\mu(\omega,k)$ to get a series $\omega=\omega(\mu,k)$, we finally
get the series expansion for the eigenvalue $A$ using
$A=n(n-1)k^2\omega$. We have checked it indeed gives the expansion (\ref{lameeigen2}).

\section{Extension to other asymptotic expansion regions}\label{sec6}

\subsection{Magnetic expansion}\label{sec6.1}

The last task we have to do is extending the asymptotic expansion in
the electric region to the magnetic and dyonic regions, as what we
have done for the Mathieu equation in Ref.~\onlinecite{wh1006b}. We stress that the
asymptotic expansion in the magnetic region has been worked out in
the literature, using purely mathematical technique. The explicit
formula is given in the book by H. J. W.
M\"{u}ller-Kirsten \cite{MullerKirstenQUANT}, which cites results from earlier original paper of E. L. Ince,
and paper of H. J. W. M\"{u}ller. It is also presented in Chapter 29 of Ref.~\onlinecite{NIST}. It expands as
\beq A&=&-i2\kappa\mu-\f{1}{2^3}(1+k^2)(4\mu^2-1)-\f{i}{2^5\kappa}[(1+k^2)^2(4\mu^3-3\mu)-4k^2(4\mu^3-5\mu)]\nn\\
&\quad&+\f{1}{2^{10}\kappa^2}(1+k^2)(1-k^2)^2(80\mu^4-136\mu^2+9)+\cdots\label{lameeigen4}\eeq
The expansion is valid for $nk\gg1, nk\gg\mu$. Note $k\sim q^{\f{1}{4}}$. We remind the
readers about the notation difference, the Floquet exponent in Ref.~\onlinecite{MullerKirstenQUANT} is denoted by $q$,
its relation to our $\mu$ is $q=2i\mu$. Also there the eigenvalue is denoted
by $\Lambda$, related to ours by $\Lambda=-A$. In the limit
$n\to\infty, k\to0$ while keep $\kappa^2=n(n-1)k^2\to-4h$ fixed, the
eigenvalue (\ref{lameeigen4}) reduces to the corresponding Mathieu
eigenvalue (\ref{mathieueigen2}), $-A\to\lambda-2h$.

Now, let us look at how to extend the Lam\'{e} eigenvalue
(\ref{lameeigen2}) in the electric region to the magnetic region.
The tool is the differential operators obtained in Sec. \ref{sec5}, and apply
them to the $\beta$-contour integral of $p_0$. In the magnetic region located
near $A\sim0$, where $\omega\ll1$, we can expand $\sqrt{\omega+\sin^2\varphi}$ as
\be
\sqrt{\omega+\sin^2\varphi}=\sin\varphi+\f{\omega}{2}\f{1}{\sin\varphi}
-\f{\omega^2}{8}\f{1}{\sin^3\varphi}+\f{\omega^3}{16}\f{1}{\sin^5\varphi}+\cdots.\ee
All the integrals now we need to do are\be
\oint_\beta\f{1}{\Delta\sin^{2n-1}\varphi}d\varphi=2\int_{\f{\pi}{2}}^{\varphi_0}\f{1}{\Delta\sin^{2n-1}\varphi}d\varphi,\qquad
n=0,1,2,3,\cdots,\ee
with $\sin\varphi_0=\f{1}{k}$, note that both $\varphi_0$ and $k$ are complex quantities. They are also given in Chapter 2 of Ref.~\onlinecite{tableofintegrals}. The first few orders of the integral
$\oint_\beta p_0d\varkappa$ give\be
\oint_\beta\sqrt{\omega+\sn^2\varkappa}d\varkappa=i\pi(\f{1}{2}\omega-\f{1+k^2}{16}\omega^2+\f{3k^4+2k^2+3}{128}\omega^3
+\cdots).\label{p0mag}\ee
Using the differential operators obtained in Sec. \ref{sec5}, we can generate
integrals $\oint_\beta p_nd\varkappa$ for $n\geqslant2$.

For the Floquet exponent in the magnetic region, our claim is \be
\mu=\f{1}{i\pi}\oint_\beta d\varkappa p(\varkappa ).\label{monomag}\ee There is a
point a bit puzzling. When we go through the circle $\beta$,
the coordinate $\varkappa$ shifts $2iK^{'}=2iK(k^{'})$,
according to the Floquet theory the exponent is defined by the relation $2iK(k^{'})\mu=\oint_\beta pd\varkappa$. Therefore
the phase $\oint_\beta pd\varkappa$ should be divided by $2iK^{'}$
to get the exponent $\mu$. But we find the correct number to divide by
is $i\pi$, in order to reproduce the known expansion (\ref{lameeigen4}).

The remaining work is the same as that in Sec. \ref{sec5}: use the
differential operators $D_{2l}(\omega,\p_\omega,k)$ to
generate higher order integrals for $p_2,p_4, \cdots$, so we get the
series expansion of $\mu=\mu(\omega,k)$. Then inverse the series
$\mu(\omega,k)$ to get a series $\omega=\omega(\mu,k)$. Finally  we
get the series expansion for the eigenvalue $A$ using
$A=n(n-1)k^2\omega=\kappa^2\omega$. With the algorithm in Appendix \ref{append4},
we have reproduced the expansion (\ref{lameeigen4}) up to higher order.

From the magnetic expansion $A(\mu,k)$ given above, use the
relation (\ref{ABrela}) and change parameters $\mu, k$ to $\nu, q$, we get
the magnetic expansion of $B(\nu,q)$. Of course, $B(\nu,q)$ is not
economic for the magnetic expansion.

\subsection{Dyonic expansion}

The eigenvalue at the dyonic point is mirror to the one at the
magnetic point. This is most obvious for $B(\nu,q)\sim\tilde{u}$.
From Fig. \ref{fig3} we know that the magnetic point
$\tilde{u}=e_2m^2/8$ is mapped to the dyonic point
$\tilde{u}=e_3m^2/8$ by $e_2\to e_3$, or
$q^{\f{1}{2}}\to-q^{\f{1}{2}}$. Also from the story of Mathieu
equation \cite{wh1006a}, we know that the exponent of magnetic region is
mapped to the exponent of dyonic region by $\nu\to i\nu$. Therefore by
the mirror map $q^{\f{1}{2}}\to-q^{\f{1}{2}}, \nu\to i\nu $ we get
the dyonic expansion $B_d$ from its magnetic expansion $B_m$.

It is also interesting to look at how the dyonic expansion of $A$ can be obtained from its magnetic expansion. For this purpose, we need to
know how the other parameters are changed by the map. It is simple
to see that under $q^{\f{1}{2}}\to-q^{\f{1}{2}}$, we have $e_2$ and
$e_3$ interchanged while $e_1$ unchanged. Furthermore, since \be
k=4q^{\f{1}{4}}\prod_{n=1}^{\infty}(\f{1+q^n}{1+q^{n-\f{1}{2}}})^4,\qquad
k^{'}=\prod_{n=1}^{\infty}(\f{1-q^{n-\f{1}{2}}}{1+q^{n-\f{1}{2}}})^4,\label{k-q}\ee
the map leads to a simple change for $k$ and $k^{'}$,
\be
k\to i\f{k}{k^{'}}, \qquad k^{'}\to\f{1}{k^{'}}.\ee
For the magnetic expansion, $A$ and $B$ satisfy \beq
B_m&=&(e_1-e_2)A_m(\mu,k)-e_2n(n-1)\nn\\
\quad&=&(e_1-e_2)A_m(\f{\nu}{\sqrt{e_1-e_2}},k(q))-e_2n(n-1).\eeq
The mirror map changes the relation to \be
B_d=(e_1-e_3)A_m(\f{i\nu}{\sqrt{e_1-e_3}},i\f{k}{k^{'}})-e_3n(n-1).\ee
The subscripts $m, d$ are used to emphasize that $A, B$ should be understood as asymptotic series in magnetic and dyonic regions.
Then substituent this into the dyonic relation
$B_d=(e_1-e_2)A_d(\mu,k)-e_2n(n-1)$, we get\beq
A_d(\mu,k)&=&\f{e_1-e_3}{e_1-e_2}A_m(\f{i\nu}{\sqrt{e_1-e_3}},
i\f{k}{k^{'}})-\f{e_3-e_2}{e_1-e_2}n(n-1)\nn\\
\quad&=&\f{e_1-e_3}{e_1-e_2}A(i\mu\sqrt{\f{e_1-e_2}{e_1-e_3}},
i\f{k}{k^{'}})-\f{e_3-e_2}{e_1-e_2}n(n-1),\eeq where the function
$A$ is the magnetic expansion (\ref{lameeigen4}) given in Subection \ref{sec6.1}. Take into account the relation of $k^2$ and $e_i$ in (\ref{k2ei}), we obtain \be
A_d=k^{'2}A(\f{i\mu}{k^{'}},
\f{ik}{k^{'}})-k^2n(n-1).\label{lameeigen5}\ee The first few
terms expand as\beq
A_d&=&-\kappa^2+i2\kappa\mu+\f{1}{2^3}(1-2k^2)(\f{4\mu^2}{k^{'2}}+1)\nn\\
&\quad&+\f{i}{2^5\kappa}[\f{(1-2k^2)^2}{k^{'}}(\f{4\mu^3}{k^{'3}}+\f{3\mu}{k^{'}})+4k^2k^{'}(\f{4\mu^3}{k^{'3}}+\f{5\mu}{k^{'}})]+\cdots.\label{lameeigen5b}\eeq
In the decoupling limit $A_d$ reduces to the corresponding Mathieu eigenvalue (\ref{mathieueigen22}), $-A_d\to\lambda-2h$.

We can perform the WKB analysis to confirm the above argument.
Set $A=-\kappa^2+\widetilde{A}$, i.e., $\omega=-1+\tomega$, then the
Lam\'{e} equation (\ref{lame3}) becomes \be
\f{\epsilon^2}{2}\Phi^{''}-(\tomega-\cn^2\varkappa)\Phi=0.\label{lame4}\ee
The WKB analysis for this equation is very similar as that in the
magnetic case, now with the leading order WKB integrand $\tp_0=i\sqrt{2}\sqrt{\tomega-\cn^2\varkappa}$, and the integral
contour is $\gamma=\alpha+\beta$. Using the amplitude variable, the integration is
\be (i\sqrt{2})^{-1}\oint_\gamma
\tp_0d\varkappa=2\int_{0}^{K+iK^{'}}\sqrt{\tomega-\cn^2\varkappa}d\varkappa
=2\int_{0}^{\varphi_0}\sqrt{\f{\tomega-\cos^2\varphi}{1-k^2\sin^2\varphi}}d\varphi.\ee
In the dyonic region, the eigenvalue is expanded for
$\kappa\gg1,\kappa\gg\mu$, i.e. $\tomega\ll1$, therefore using
\be
\sqrt{\tomega-\cos^2\varphi}=
i(\cos\varphi-\f{\tomega}{2}\f{1}{\cos\varphi}-\f{\tomega^2}{8}\f{1}{\cos^3\varphi}-\f{\tomega^3}{16}\f{1}{\cos^5\varphi}+\cdots),\ee
the integrals we need to do are \be
\oint_\gamma\f{1}{\Delta\cos^{2n-1}\varphi}d\varphi=2\int_{0}^{\varphi_0}\f{1}{\Delta\cos^{2n-1}\varphi}d\varphi,\qquad
n=0,1,2,3,\cdots.\ee The first few of them give\be
\oint_\gamma\sqrt{\tomega-\cn^2\varkappa}d\varkappa=
-\pi(\f{1}{2k^{'}}\tomega+\f{1-2k^2}{16k^{'3}}\tomega^2+\f{8k^4-8k^2+3}{128k^{'5}}\tomega^3+\cdots).\label{p0dyon}\ee

For the higher order WKB integrals, they are given by the differential operators $\widetilde{D}_n(\tomega,\p_{\tomega},k)$ acting on the leading order integral.
As in Subsection \ref{sec6.1}, for odd order integrals $\oint_\gamma\tp_{2l+1}d\varkappa=0$, therefore we set $\widetilde{D}_{2l+1}(\tomega,\p_{\tomega},k)=0$. For even order integrals, we get nontrivial $\widetilde{D}_{2l}(\tomega,\p_{\tomega},k)$, for example $\widetilde{D}_2(\tomega,\p_{\tomega},k)$ is obtained from \beq \oint_\gamma
\tp_2d\varkappa&=&-\f{1}{12}[(-1+k^2+2\tomega-4k^2\tomega+3k^2\tomega^2)\p_{\tomega}^{2}\nn\\
&\quad&+(1-2k^2+3k^2\tomega)\p_{\tomega}-\f{3}{4}k^2]\oint_\gamma
\tp_0d\varkappa.\label{p2lamedyon}\eeq
The  Appendix \ref{append4} gives the algorithm to obtain all $\widetilde{D}_{2l}(\tomega,\p_{\tomega},k)$, in a simpler way they can also be obtained from $D_{2l}(\omega,\p_\omega,k)$ using a property (\ref{Delec2Ddyon}) explained below.

We find that for the dyonic expansion, the Floquet exponent is given by
\be \mu=\f{k^{'}}{\pi}\oint_\gamma d\varkappa \tp(\varkappa).\label{monodyon}\ee  The factor
$k^{'}/\pi$ also does not follow the conventional definition of the Floquet exponent,
at the moment we do not have a satisfying explanation for this, but the relation is indeed satisfied. We check this relation against (\ref{lameeigen5}) up to high order.

Some interesting phenomena are observed and deserve mentioned.
Look at the leading order contour integrals for the magnetic and dyonic expansions,
formulae (\ref{p0mag}) and (\ref{p0dyon}), they are related in a
simple way, \be k\to i\f{k}{k^{'}},\quad
\omega\to-\tomega,\qquad\m{leads\quad to}\qquad
\f{1}{\epsilon}\oint_\beta\sqrt{\omega+\sn^2\varkappa}d\varkappa\to\f{-1}{\epsilon}\oint_\gamma\sqrt{\tomega-\cn^2\varkappa}d\varkappa.\ee
Note that $\epsilon\sim\kappa^{-1}\sim k^{-1}$ also changes as
$\epsilon\to-ik^{'}\epsilon$. Then look at the differential
operators, formulae (\ref{p2lame}) and
(\ref{p2lamedyon}), they are also related in a
simple way, \be k\to i\f{k}{k^{'}},\quad
\omega\to-\tomega,\qquad\m{leads}\quad\m{to}\qquad
\epsilon^nD_n(\omega,\p_\omega,k)\to
\epsilon^n\widetilde{D}_n(\tomega,\p_{\tomega},k),\quad
n=2.\label{Delec2Ddyon}\ee The relation continues to hold for larger $n$. Combining the two facts,
under the mirror map we have\be \oint_\beta d\varkappa
p(\varkappa)\to-\oint_\gamma d\varkappa
\tp(\varkappa),\qquad\m{i.e.}\qquad \mu\to i\f{\mu}{k^{'}}.\ee This
is consistent with (\ref{lameeigen5}) where the Floquet exponent appears as
$i\mu/k^{'}$.

\section{Conclusion}\label{sec7}

Combinatorics is a very interesting and fruitful subject of
mathematics, its relation to integrable models is not
new \cite{Zinn-Justin}. In recent years, some progresses continue to
reveal its fascinating connection to quantum field theory,
integrable model, and string/M theory \cite{no0306, NS0908, nekr2009,
agt}. Based on a relation between N=2 gauge theory and
quantum integrable system \cite{NS0908}, we provide a quantum
field theory approach to spectral problem of the Mathieu equation and the Lam\'{e}
equation. The approach is combinatorial, based on the instanton
counting method \cite{instcount}. We use the electro-magnetic duality in gauge theory to derive all three asymptotic expansions of eigenvalue.

Our main results are about the gauge theory explanation of the asymptotic expansion formulae (\ref{lameeigen}), (\ref{lameeigen4}) and
(\ref{lameeigen5}) for the eigenvalue of the Lam\'{e} equation, the discussion relies on the Floquet theory of periodic differential equation.
While the expansions (\ref{lameeigen}), (\ref{lameeigen4}) were obtained by some other independent methods,
to our knowledge the expansion (\ref{lameeigen5}) is a new result. E. Langmann's paper Ref.~\onlinecite{Langmann2004b} gives an expansion equivalent
to (\ref{lameeigen}), we have shown the equivalence in  Appendix \ref{append3}.
We perform the WKB analysis in Sections \ref{sec5} and \ref{sec6}, given all three expansions. Also in the decoupling limit the
expansions correctly reduce to expansions of the Mathieu eigenvalue.

In this paper we do not discuss the combinatorial connection to the
eigenfunctions: the Mathieu function and Lam\'{e}
function/polynomial. There are examples for some integrable models a full combinatorial
approach to the eigenvalue and eigenstate exists. The
Calogero-Sutherland model is a famous example, see a recent
discussion in Ref.~\onlinecite{wxy1107}, and relevant references therein. In
fact, the Calogero-Sutherland model is the trigonometric limit of
the elliptic Calogero-Moser model. It is argued that in the context of Alday-Gaiotto-Tachikawa(AGT) relation \cite{agt},
the instanton partition function of gauge theory with full surface operator, in the semiclassical limit $\epsilon_2\to0$, gives
the eigenfunction of the corresponding quantum integrable model \cite{at1005, mt1006}. This maybe helpful for an explicit combinatorial
construction of the eigenfunction.

While the basic tools of the paper, such as the Seiberg-Witten theory,
instanton calculation, have been extensively studied by mathematical
physicists and have a solid mathematical foundation, some points of our results lack
rigorous mathematical proof. This includes the precise meaning of the potential
$\mathcal {F}$ for the differential equations, the relation between the instanton method and Langmann's method.
Another unsatisfying point is about the puzzling factors $(i\pi)^{-1}$ and $k^{'}(\pi)^{-1}$ in the relations (\ref{monomag}) and (\ref{monodyon}).
In our paper Ref.~\onlinecite{wh2014} we use a different method to study the asymptotic expansions for the eigenvalue of Schr\"{o}dinger operator with elliptic potentials, including a new example of the ellipsoidal wave equation,
find the same factors are needed to define the Floquet exponent in order to produce asymptotic expansions consistent with known results.
These results suggest that the Floquet theory for doubly-periodic elliptic potentials requires a new treatment unrecorded in conventional Floquet theory.
Therefore, we do not present a mathematically rigorous treatment of the Mathieu and the Lam\'{e} equations in this single paper,
but just to show how the quantum field theories are related to these classical differential
equations and can be used to obtain some consistent results of them. We hope to clarify some of the points in the future.

\section*{Acknowledgments}

During the progress of this work I was supported by NSFC No. 11031005, through ZJU, China;
later I was supported by FAPESP No. 2011/21812-8, through IFT-UNESP, Brazil.
I thank an anonymous referee for bringing Langmann's results Ref.~\onlinecite{Langmann2004b} to my attention and suggestions for better presentation of the paper.

\appendix

\section{Modular functions, Theta constants\label{append1}}

We collect some basic facts about modular functions we have used in
this paper. Our main references are Refs.~\onlinecite{Wang-Guo, PanPan, AEmodular}.

The complete elliptic integrals of the first kind is defined by
\be
K=\int_0^{\f{\pi}{2}}\f{d\varphi}{\sqrt{1-k^2\sin^2\varphi}}=\f{\pi}{2}F(\f{1}{2},\f{1}{2},1;k^2),\ee
where $F(a,b,c;x)$ is the hypergeometric function. The complete elliptic integrals of the
second kind is defined by \be
E=\int_0^{\f{\pi}{2}}d\varphi\sqrt{1-k^2\sin^2\varphi}=\f{\pi}{2}F(-\f{1}{2},\f{1}{2},1;k^2).\ee

Then we define the elliptic nome $q=e^{2\pi i\tau}$ through $\tau=\f{iK^{'}}{K}$, where $K^{'}=K(k^{'})$. The theta constant
$\theta_i$, which are the theta function $\theta_i(z;q)$ at $z=0$,
are given by \be
\begin{aligned}
&\theta_1(q)=0,\qquad \theta_2(q)=2q^{\f{1}{8}}\sum_{n=0}^{\infty}q^{\f{n(n+1)}{2}},\\
&\theta_3(q)=1+2\sum_{n=1}^{\infty}q^{\f{n^2}{2}},\qquad \theta_4(q)=1+2\sum_{n=1}^{\infty}(-1)^nq^{\f{n^2}{2}},
\end{aligned}
\ee they satisfy the relation $\theta_3^4=\theta_2^4+\theta_4^4$.

From the theta constant we define
\be
e_1=\f{2}{3}\theta_3^4-\f{1}{3}\theta_2^4,\qquad
e_2=-\f{1}{3}(\theta_3^4+\theta_2^4),\qquad
e_3=-\f{1}{3}\theta_3^4+\f{2}{3}\theta_2^4.
\ee
$e_i$ satisfy $e_1+e_2+e_3=0$, they are related to the
Weierstrass function, the roots $z_i$ of
$\wp^{'2}(z)=4\wp^3(z)-g_2\wp(z)-g_3=0$ satisfy $\wp(z_i)=e_i$.
We have the following relation between $k$ and $q$, \be
k^2=\f{e_3-e_2}{e_1-e_2}=\f{\theta_2^4}{\theta_3^4},\label{k2ei}\ee which is
equivalent to the relation in (\ref{kqrelat}).

The second Eisenstein series is represented by \beq
E_2(q)&=&1-24\sum_{n=1}^{\infty}\f{nq^n}{1-q^n}=1-24\sum_{n=1}^{\infty}\f{q^n}{(1-q^n)^2}\nn\\
\quad&=&1-24q-72q^2-96q^3-168q^4-144q^5-\cdots.\eeq The equality of
the two summation would be obvious from the following, \beq
\sum_{n=1}^{\infty}\f{nq^n}{1-q^n}&=&\sum_{n=1}^{\infty}\sum_{m=0}^{\infty}nq^{n(m+1)}=\sum_{n=1}^{\infty}\sum_{m=1}^{\infty}nq^{nm},\nn\\
\sum_{n=1}^{\infty}\f{q^n}{(1-q^n)^2}&=&\sum_{n=1}^{\infty}(\f{q^n}{(1-q^n)}\f{1}{(1-q^n)})
=\sum_{n=1}^{\infty}(\sum_{r=1}^{\infty}q^{rn})(\sum_{s=0}^{\infty}q^{sn})\nn\\
\quad&=&\sum_{n=1}^{\infty}(\sum_{r=1}^{\infty}\sum_{s=0}^{\infty}q^{(r+s)n})=\sum_{n=1}^{\infty}\sum_{t=1}^{\infty}tq^{tn}.\eeq
The last step uses the fact for a fixed $t=r+s$, there are exactly
$t$ pairs of $(r\geqslant1,s\geqslant0)$ satisfy it. $E_2$ is related to the
Dedekind Eta function by $\f{d}{d\tau}\ln\eta(\tau)=\f{i\pi}{12}E_2(\tau)$.

\section{Compare to Langmann's expansion\label{append3}}

E. Langmann developed an algorithm to derive the eigenvalue and
eigenfunction of elliptic Calogero-Moser(-Sutherland) model for
general particle number \cite{Langmann2004a, Langmann2004b}. In his
paper Ref.~\onlinecite{Langmann2004b} a series expansion for the 2-body eigenvalue
when $q\ll1$ is given(formulae 26a-d). Although his method is
different from the instanton and WKB methods we present in this paper,
we will show that his expansion is equivalent to our expansion
(\ref{lameeigen}). Langmann considered the 2-particle Shr\"{o}dinger
equation, \be [-(\f{\p^2}{\p x_1^2}+\f{\p^2}{\p
x_2^2})+2\lambda(\lambda-1)V(x_1-x_2)]\psi(\vec{x})=\mathcal
{E}\psi(\vec{x}).\ee with potential $V(x)=-\wp(x)+c_0$, \beq
c_0&=&\f{1}{12}-2\sum_{m=1}^{\infty}\f{1}{(q^{m/2}-q^{-m/2})^2}=\f{1}{12}E_2(q)\nn\\
\quad&=&\f{1}{12}-2q-6q^2-8q^3-14q^4+\cdots.\eeq After changing
variables, the equation becomes the Lam\'{e} equation (\ref{lame2})
describing the relative motion of particles. The eigenvalue $\mathcal {E}$ is
explicitly given as an expansion in $q$:
$\mathcal{E}=\mathcal{E}_0+\mathcal{E}_1q+\mathcal{E}_2q^2+\mathcal{E}_3q^3+\cdots$.
Especially notice that, in the notation of paper Ref.~\onlinecite{Langmann2004b},
the relation \be 2\mathcal{E}_0=P^2+(n_1+n_2)^2,\ee the term
$(n_1+n_2)^2$ is the kinetic energy of center-of-mass motion.

The relation between the eigenvalues $\mathcal {E}$ and our $B$ is\be
B=-2\mathcal{E}+(n_1+n_2)^2+4\lambda(\lambda-1)c_0,\label{EBrelation}\ee
provided we identify parameters as $P=\nu, \lambda=n$. Then using the
relation (\ref{EBrelation}) to  Langmann's expansion of $\mathcal
{E}$, we can rewrite our expansion (\ref{lameeigen}) in a slightly
different form
\beq B&=&-\nu^2+\f{n(n-1)}{3}E_2(q)-\f{8n^2(n-1)^2q}{\nu^2-1}\nn\\
&\quad&-\f{8n^2(n-1)^2q^2}{(\nu^2-1)^3(\nu^2-4)}\times[n^2(n-1)^2(5\nu^2+7)-12n(n-1)(\nu^2-1)^2\nn\\
&\quad&+6(\nu^2-1)^2(\nu^2-2)]+\mathcal{O}(q^3).\label{lameeigen3}\eeq
Compare (\ref{lameeigen}) and (\ref{lameeigen3}), the difference
between them is that we have collected some terms coming from
instanton contribution in (\ref{lameeigen}) which summed into $\f{1}{3}n(n-1)(1-E_2)$ , together with
$-\f{1}{3}n(n-1)(1-2E_2(q))$, gives the term $\f{1}{3}n(n-1)E_2(q)$ in (\ref{lameeigen3}).
This fact can be checked to higher order by calculating higher order instanton effects in
gauge theory. Note that the value $\f{1}{3}n(n-1)(1-E_2)$ exactly corresponds to the extra U(1) term contained in the prepotential of U(2) gauge theory, see subsection \ref{subsec4.2}.

\section{A method to obtain the differential operators\label{append4}}

When we derive the differential operators $D_n(\omega,\p_\omega,k)$,
we need to separate from the integrand some terms of total derivative.
By discarding these total derivative terms from the contour integral,
we can simplify the differential operators as far as possible.
It is necessary to find a systematic method to deal with higher order WKB perturbations to make the program complete.
After some trial and error, we find a workable method.

The method is carried out as the following.

{\bf The first step}, we solve the WKB relation for $p_n(\varkappa)$,
explicitly carry out the derivatives and the final expressions are of the form,
\be p_{2l+1}(\varkappa)=\f{\sn\varkappa\cn\varkappa\dn\varkappa}{(\omega+\sn^2\varkappa)^{3l+1}}\times(\m{polynomial of }\sn^2\varkappa),\qquad l=0,1,2,\cdots,\ee
\be p_{2l}(\varkappa)=\f{1}{(\omega+\sn^2\varkappa)^{3l-\f{1}{2}}}\times(\m{polynomial of }\sn^2\varkappa),\qquad l=0,1,2,\cdots.\ee

The $p_{2l+1}(\varkappa)$ can be written as a total derivative term, so its contour integral is zero.
$p_1\sim\p_\varkappa(\ln p_0)$ is special. The polynomials in the numerator of $p_{2l+1}$ for $l\geqslant1$ are of the form $c_*\sn^{6l-2}\varkappa+c_*\sn^{6l-4}\varkappa+\cdots+c_*\sn^{2}\varkappa+c_*$, where we use  $c_*$ to represent all coefficients whose explicit form is not important here, they are some polynomials of $k^2, \omega$. Keep in mind in the following formulae all $c_*$ (and later $a_*$) could be different from each other.
Set $\omega+\sn^2\varkappa=t$, then $p_{2l+1}$ can be written as
\be p_{2l+1}(\varkappa)=(\sn\varkappa\cn\varkappa\dn\varkappa)(c_*t^{-(3l+1)}+c_*t^{-3l}+c_*t^{-(3l-1)}+\cdots c_*t^{-3}+c_*t^{-2}).\label{piexpandodd}\ee
Every term in (\ref{piexpandodd}) is a total derivative given by
\be c_*\p_{\varkappa}(\omega+\sn^2\varkappa)^{-l^{'}}, \qquad l^{'}=1,2,\cdots,3l;\qquad l\geqslant1,\ee
with proper choice of the coefficient. Therefore, $\oint d\varkappa p_{2l+1}(\varkappa)=0$ as expected.

So we only need to deal with $p_{2l}$ for $l\geqslant0$. The polynomials in the numerator of $p_{2l}$ are of the form $c_*\sn^{6l}\varkappa+c_*\sn^{6l-2}\varkappa+\cdots+c_*\sn^{2}\varkappa+c_*$. Then $p_{2l}$ can be expanded as a polynomial of $t$,
let us denote it as $p_{2l}^{(I)}$,
\be p_{2l}^{(I)}(\varkappa)=c_*t^{-(3l-\f{1}{2})}+c_*t^{-(3l-\f{3}{2})}+c_*t^{-(3l-\f{5}{2})}+\cdots c_*t^{-\f{1}{2}}+c_*t^{\f{1}{2}}.\label{piexpand}\ee
It is generated by a differential operator acting on $p_0\sim\sqrt{t}$,
\be \oint d\varkappa p_{2l}^{(I)}(\varkappa)=(c_*\p_{\omega}^{3l}+c_*\p_{\omega}^{3l-1}+c_*\p_{\omega}^{3l-2}+\cdots c_*\p_{\omega}+c_*)\oint d\varkappa p_0(\varkappa).\ee

{\bf The second step} is to reduce the order of the differential operator for $p_{2l}^{(I)}$. The method is to find proper total derivative terms like $\p_\varkappa(***)$ that can substitute terms of  $c_*t^{-(3l-\f{1}{2})}+c_*t^{-(3l-\f{3}{2})}+\cdots+c_*t^{-(2l+\f{1}{2})}$ in (\ref{piexpand}), then abandon them in the contour integral $\oint p_{2l}^{(I)}d\varkappa$.
Therefore, the lowest order of the polynomial integrand (\ref{piexpand})  is increased and the corresponding differential operator for $p_{2l}^{(I)}$ is also simplified. We find the total derivative terms are generated by
\be c_*\p^2_{\varkappa}(\omega+\sn^2\varkappa)^{-(3l-\f{3}{2}-l^{'})}, \qquad l^{'}=1,2,\cdots,2l-1.\label{totalDterm}\ee A nice property of this term is that its final expression can be expressed in terms of $\sn^2\varkappa$, or in terms of $t$, as
\be c_*t^{-(3l+\f{1}{2}-l^{'})}+c_*t^{-(3l-\f{1}{2}-l^{'})}+c_*t^{-(3l-\f{3}{2}-l^{'})}+c_*t^{-(3l-\f{5}{2}-l^{'})}.\label{totalDexpand}\ee

Now we can discard some terms in (\ref{piexpand}) using (\ref{totalDterm}). For $l^{'}=1$, choosing proper coefficient in (\ref{totalDterm}),
we can make the coefficient of the first term in (\ref{totalDexpand}) equals to the coefficient of the first term in (\ref{piexpand}),
hence
\beq p_{2l}^{(I)}(\varkappa)&=&\left[c_*\p_{\varkappa}\left(\f{\sn\varkappa\cn\varkappa\dn\varkappa}{(\omega+\sn^2\varkappa)^{3l-\f{3}{2}}}\right)
-\left(c_*t^{-(3l-\f{3}{2})}+c_*t^{-(3l-\f{5}{2})}+c_*t^{-(3l-\f{7}{2})}\right)\right]\nn\\
&\quad&+\left(c_*t^{-(3l-\f{3}{2})}+c_*t^{-(3l-\f{5}{2})}+\cdots c_*t^{-\f{1}{2}}+c_*t^{\f{1}{2}}\right).\eeq
Now the total derivative term can be abandoned, and the order of $p_{2l}^{(I)}$ is increased to $t^{-(3l-\f{3}{2})}$, meanwhile the coefficients of other three terms are changed. Repeat this process for $l^{'}=2$, we can increase the order of $p_{2l}^{(I)}$ to $t^{-(3l-\f{5}{2})}$, an so on. For every $l^{'}\in\lbrace1,2,\cdots,l\rbrace$, the process of removing (\ref{totalDterm}) is carried out once. After the process we get the $p_{2l}^{(II)}$,
\be p_{2l}^{(II)}(\varkappa)=c_*t^{-(2l-\f{1}{2})}+c_*t^{-(2l-\f{3}{2})}+c_*t^{-(2l-\f{5}{2})}+\cdots c_*t^{-\f{1}{2}}+c_*t^{\f{1}{2}},\ee
and the associated differential operator,
\be \oint d\varkappa p_{2l}^{(II)}(\varkappa)=(c_*\p_{\omega}^{2l}+c_*\p_{\omega}^{2l-1}+c_*\p_{\omega}^{2l-2}+\cdots c_*\p_{\omega}+c_*)\oint d\varkappa p_0(\varkappa).\label{pifromp0}\ee

{\bf The third step} is to minimize the differential operator. After carrying out the process for $l^{'}=1,2,\cdots,l$, we get $p_{2l}^{(II)}$ and the corresponding differential operator of the form in (\ref{pifromp0}).
Remember that the coefficients $c_*$ in (\ref{pifromp0}) are polynomials of $k^2,\omega$, it turns out that these coefficients can be further simplified. The reason is the following: we assume the differential operators for the Mathieu equation, conjectured in Ref.~\onlinecite{wh1006b}, take the simplest form we can get through the simplification process. We call these differential operators {\em minimal}.
The differential operators of the Lam\'{e} equation, if they are minimal, should reduce to the minimal differential operators of the Mathieu equation in the limit
$k\to0, \omega\to (w-1)/2$. However, we find the differential operator obtained after performing the second step for $l^{'}=1,2,\cdots,l$ is not minimal,
some redundant terms can be further discarded. We should continue the process for $l^{'}=l+1,l+2,\cdots,2l-1$ to remove total derivative terms of (\ref{totalDterm}).
For every $l^{'}\in\lbrace l+1,l+2,\cdots,2l-1\rbrace$, the process of removing (\ref{totalDterm}) may be repeated for a few times, depending on $l$.
The simplification does not further decrease the order of the differential operator in (\ref{pifromp0}), but simplifies its coefficients to their minimal form.

Some details about this step would be helpful. Unlike the previous step where the whole coefficient $c_*$ is subtracted,
the new issue here is that only some terms in the coefficients $c_*$ should be subtracted,
we have to determine which part. The key point here is that the criterion of ``minimal" comes from results of the Mathieu equation,
so in order to get a clue,  in this step we need to track the limit $k\to0, \omega\to(w-1)/2$ for the formulae we have obtained,
and compare them to the results of the Mathieu equation.
In this limit the $p_{2l}^{(II)}$ obtained from the second step becomes
\beq \underrightarrow{\small{\mbox{lim}}}p_{2l}^{(II)}&=&(\cdots+a_*w^{l-4}+a_*w^{l-2}+a_*w^l)\hat{t}^{-(2l-\f{1}{2})}\nn\\
&\quad&+(\cdots+a_*w^{l-5}+a_*w^{l-3}+a_*w^{l-1})\hat{t}^{-(2l-\f{3}{2})}+\cdots+a_*w\hat{t}^{-(l+\f{1}{2})}+a_*\hat{t}^{-(l-\f{1}{2})}, \label{p2llim}\eeq
with $a_*$ some numerical coefficients, the powers of $w$ are non-negative, and $\hat{t}\sim\sqrt{w-\cos2\varkappa}$ is the limit of $t$.
From the results of the Mathieu equation \cite{wh1006b}, we know that the minimal form of the expression on the right hand side should be
$a_*w^l\hat{t}^{-(2l-\f{1}{2})}+a_*w^{l-1}\hat{t}^{-(2l-\f{3}{2})}+\cdots+a_*w\hat{t}^{-(l+\f{1}{2})}+a_*\hat{t}^{-(l-\f{1}{2})}$,
this indicates all other terms can be represented by total derivative terms and should be subtracted.
It turns out that we should use the formula (\ref{totalDterm}) with $l^{'}=l+1,l+2,\cdots,2l-1$. For example,
in order to get rid of the term $a_*w^{l-2}\hat{t}^{-(2l-\f{1}{2})}$ in (\ref{p2llim}) we should subtract a term from $p_{2l}^{(II)}$ by
\be p_{2l}^{(II)}-a_*(2\omega+1)^{l-2}\p^2_{\varkappa}(\omega+\sn^2\varkappa)^{-(3l-\f{3}{2}-(l+1))},\ee
the coefficient can be determined by comparing the coefficients of $w^{l-2}\hat{t}^{-(2l-\f{1}{2})}$ in (\ref{p2llim}) and in
$\underrightarrow{\small{\mbox{lim}}}(2\omega+1)^{l-2}\p^2_{\varkappa}(\omega+\sn^2\varkappa)^{-(3l-\f{3}{2}-(l+1))}$.
Similarly we can get rid of $w^{l-2s}\hat{t}^{-(2l-\f{1}{2})}$, with $s=2,3,\cdots[\f{l}{2}]$,
by subtracting a term made of $(2\omega+1)^{l-2s}\p^2_{\varkappa}(\omega+\sn^2\varkappa)^{-(3l-\f{3}{2}-(l+1))}$.
The formula (\ref{totalDterm}) with $l^{'}=l+1$ is repeatedly used for $\f{l}{2}$(for even $l$)  or $\f{l-1}{2}$(for odd $l$) times.
In the same way we can get rid of the remaining unnecessary terms in the second non-minimal coefficient, $a_*w^{l-3}\hat{t}^{-(2l-\f{3}{2})},a_*w^{l-5}\hat{t}^{-(2l-\f{3}{2})},\cdots$, etc,
the formula (\ref{totalDterm}) with $l^{'}=l+2$ is repeatedly used for $\f{l}{2}-1$ (for even $l$)  or $\f{l-1}{2}$ (for odd $l$) times.
Continue the process we can minimize all coefficients in (\ref{p2llim}), in general the coefficient of the term of order $\hat{t}^{-(2l-\f{1}{2}-r)}$ with $0\leqslant r\leqslant l-2$ is minimized by repeatedly using (\ref{totalDterm}) with $l^{'}=l+1+r$ for $\f{l-r}{2}$ (for even $l-r$)  or $\f{l-r-1}{2}$ (for odd $l-r$) times.

At last we get the minimal polynomial $p_{2l}^{(III)}$
and the associated minimal differential operator,
\be \oint d\varkappa p_{2l}^{(III)}(\varkappa)=(c_*\p_{\omega}^{2l}+c_*\p_{\omega}^{2l-1}+c_*\p_{\omega}^{2l-2}+\cdots c_*\p_{\omega}+c_*)\oint d\varkappa p_0(\varkappa),\ee
now with $c_*$, still polynomials of $k^2,\omega$, be the minimal coefficients.

Using this method, we successfully derive differential operators for $p_{2l}$ of the Lam\'{e} equation,
for the first few $l$. In the limit $k\to0, \omega\to(\omega-1)/2$,
these differential operators correctly reduce to the minimal differential operators of the Mathieu equation derived in Ref.~\onlinecite{wh1006b}.

Apparently, the minimal differential operators of the Lam\'{e} equation in the form (\ref{lame4}) can be obtained in the same way,
with the total derivative terms for $p_{2l}$ generated by
\be c_*\p^2_{\varkappa}(\tomega-\cn^2\varkappa)^{-(3l-\f{3}{2}-l^{'})}, \qquad l^{'}=1,2,\cdots,2l-1.\label{totalDterm2}\ee
The relation between $D_n(\omega,\p_\omega,k)$ and $\widetilde{D}_n(\tomega,\p_{\tomega},k)$ in (\ref{Delec2Ddyon}) holds for $n\geqslant2$.
And the differential operators of the Mathieu equation can be obtained following the same steps,
with the total derivative terms for $p_{2l}$ generated by
\be c_*\p^2_{z}(w-\cos2z)^{-(3l-\f{3}{2}-l^{'})}, \qquad l^{'}=1,2,\cdots,2l-1,\ee with the notation used in Ref.~\onlinecite{wh1006b}.
This also gives an explanation for the conjectural form of  $D_n(w,\p_w)$ in that paper (in this paper denoted by $D_n(u,\p_u)$).

\end{document}